\documentclass[a4paper]{article} 

\usepackage[dvips]{graphicx}
\usepackage{amssymb}

\newcommand{\argmin}{\mathop{\arg\min}\limits}

\providecommand{\keywords}[1]
{
  \small	
  \textbf{\textit{Keywords---}} #1
}

\title{
Nonlinear trend of COVID-19
\\
infection time series 
}

\author
{
Fumihiko Ishiyama
\\
\small
NTT Space Environment and Energy Labs.
\\
\small
Nippon Telegraph and Telephone Corp., 
Tokyo, 180-8585, Japan
}

\date{}

\begin{document}

\maketitle 

\begin{abstract} 
We have developed a nonlinear method of time series analysis
that allows us to obtain multiple nonlinear trends without harmonics
from a given set of numerical data.
We propose to apply the method 
to recognize the ongoing status of COVID-19 infection
with an analytical equation for nonlinear trends.
We found that there is only a single nonlinear trend,
and
this result justifies the use of a week-based infection growth rate.
In addition, 
the fit with the obtained analytical equation for the nonlinear trend
holds for a duration of more than 
three months for the Delta variant infection time series.
The fitting also visualizes the transition to the Omicron variant.
\end{abstract}


\section{Introduction}

Understanding the COVID-19 infection status,
such as ``what's going on?'' and ``what happens next?'',
has been a major issue in the last couple of years.

Various forecasting methods have been studied to address this issue.
Among them, machine learning (ML) based studies have played a major role.
For example, various methods such as 
Linear Regression (LR),
Least Absolute Shrinkage and Selection Operator (LASSO),
Support Vector Machine (SVM),
and
Exponential Smoothing (ES)
were compared to achieve better prediction,
and ES was found to have the best performance among the methods 
\cite{ieee-access}.
In addition, other methods such as 
Auto-Regressive Integrated Moving Average (ARIMA) based approach \cite{ieee-s}, 
and Gaussian Process Regression (GPR) 
based approach \cite{ieee-y} were also studied
for forecasting  COVID-19 infection.

In contrast to their focusing on ``what happens next?'', 
we focus on ``what's happening now?'' in this paper.
That is, we propose a framework to recognize nonlinear phenomena,
and apply the framework to the COVID-19 infection time series
to recognize ``what happens now''.

We have a unique nonlinear method of time series analysis
that provides an  analytical perspective on numerically obtained nonlinear time series
\cite{cspa,isspit,ieice,thesis,git},
and 
we apply the method to COVID-19 daily new cases in Japan
to obtain analytical equation for nonlinear trends.

As
there are no similar 
conventional methods, 
we show an example of  results obtained with our method in a previous work
\cite{cqg}
to illustrate our approach.

\begin{figure}[hbt]
\begin{center}
\includegraphics[width=120mm]{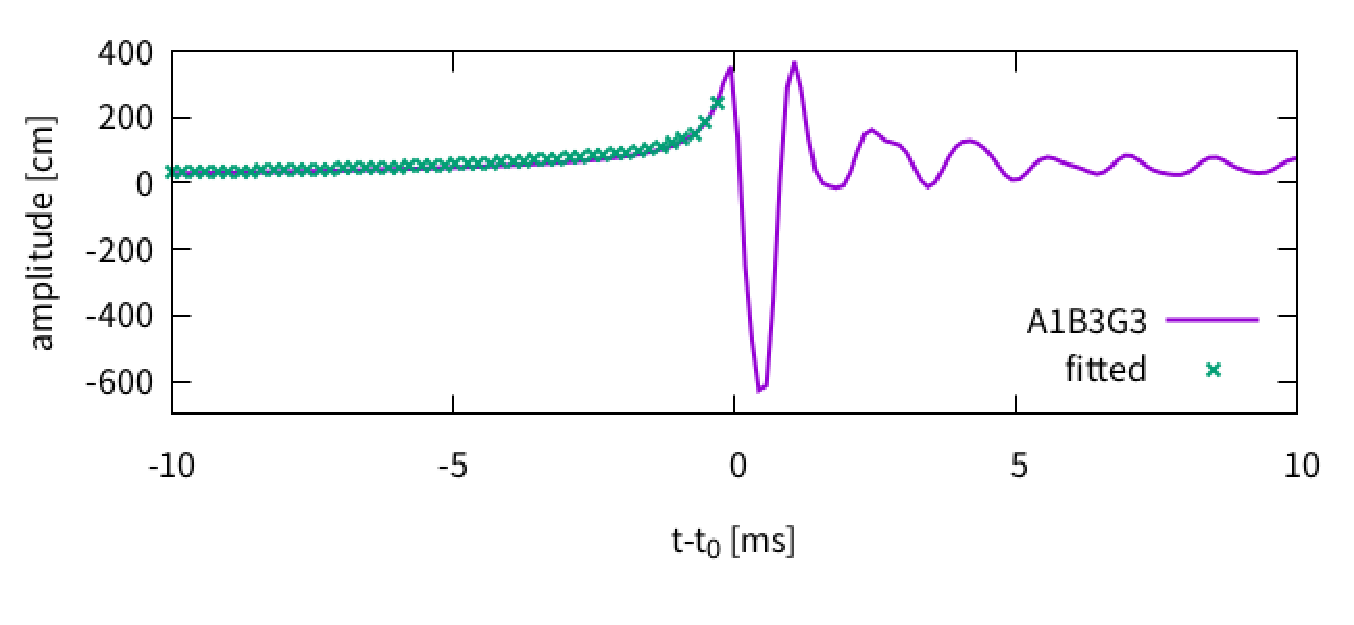}
\end{center}
\caption{
Time series of gravitational wave for analysis.
Line corresponds to Eq.~(\ref{eq-cqg-ts}) is also plotted.
}\label{fig-cqg}
\end{figure}

Figure \ref{fig-cqg} shows a numerically calculated time series $S(t)$
of a gravitational wave named A1B3G3,
which corresponds to black hole generation at $t=t_0$ \cite{dimm}.
We applied our method to the time series 
before the black hole generation
$t<t_0$,
where there are no harmonics.
Instead, there are two series of nonlinear trends with $f=0$.

\begin{figure}[hbt]
\begin{center}
\includegraphics[width=120mm]{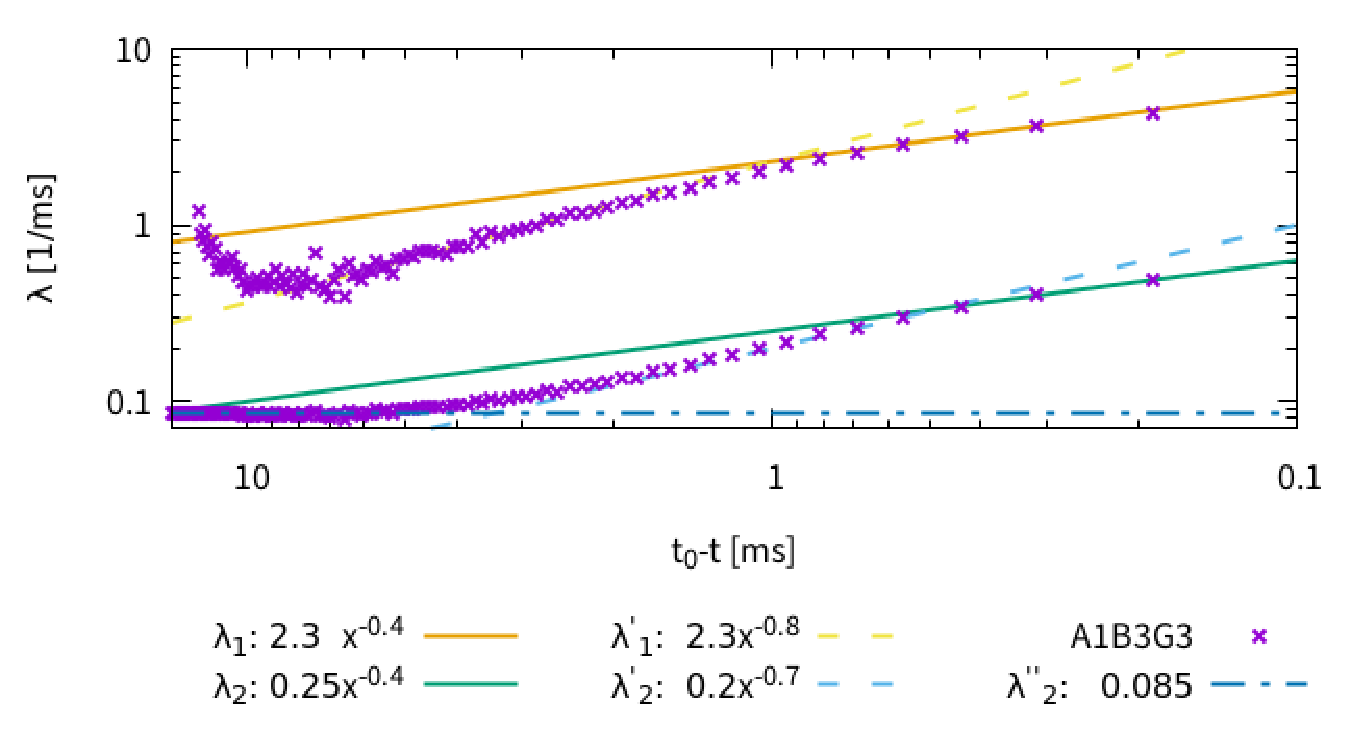}
\end{center}
\caption{
Obtained two nonlinear trends 
($\lambda_1,~\lambda^\prime_1$
and
$\lambda_2,~\lambda^\prime_2,~\lambda^{\prime\prime}_2$)
with several phase transitions.
}\label{fig-cqg-b}
\end{figure}

Time series $S(t)$ with multiple
nonlinear trends $\lambda_m(t)$ 
is expressed 
in the form of 
\begin{equation}
S(t)=\sum_{m=1}^M e^{\int \lambda_m(t) dt},
\end{equation}
and the fitted $\lambda_m(t)$ 
with our method are plotted in Fig.~\ref{fig-cqg-b}.
There are two nonlinear trends ($M=2$) with several phase transitions.

Obtained analytical equations for the nonlinear trends 
just before $t=t_0$ are
\begin{eqnarray}
\lambda_1(t)&=&2.3 (t_0-t)^{-0.4}
\\
\lambda_2(t)&=&0.25 (t_0-t)^{-0.4},
\end{eqnarray}
and the corresponding time series becomes
\begin{eqnarray}
S_0(t)
&\simeq&
702 e^{\int_{t_0}^t \lambda_1(\tau) d\tau}
+
168 e^{\int_{t_0}^t \lambda_2(\tau) d\tau} 
\nonumber
\\
&=&
702 e^{2.3 \int_{t_0}^t (t_0-\tau)^{-0.4} d\tau}
+
168 e^{0.25 \int_{t_0}^t (t_0-\tau)^{-0.4} d\tau}
\nonumber
\\
&=&
702 e^{-\frac{2.3}{0.6} (t_0-t)^{0.6}}
+
168 e^{-\frac{0.25}{0.6} (t_0-t)^{0.6}}. \label{eq-cqg-ts}
\end{eqnarray}
The fitted time series of Eq.~(\ref{eq-cqg-ts}) is plotted in Fig.~\ref{fig-cqg}.
The time series around $t_0-t \sim 2$ ms and $t_0-t > 10$~ms are
\begin{eqnarray}
S_1(t) &\simeq& 
29 e^{-\frac{2.3}{0.2} (t_0-t)^{0.2}}
+
96 e^{-\frac{0.2}{0.3} (t_0-t)^{0.3}}
\\
S_2(t) &\simeq& \hspace{28mm} 29 e^{ -0.085 (t_0-t)}
\end{eqnarray}
respectively.
Each equation corresponds to each phase.

This kind of analytical expression becomes possible with our method,
and we apply the method to a COVID-19 infection time series.

In the following sections, 
we introduce our method of nonlinear time-frequency analysis,
apply the method to the analysis of a COVID-19 infection time series, 
and finally, we conclude the paper.


\section{Method}

Our method is based on a mode decomposition 
with general complex functions, 
which organize nonlinear oscillators 
\cite{cspa,isspit,ieice,thesis,git},
and we calculate the local linearized solution of the decomposition.

Nonlinear trends extraction is a part of our method.
Obtained nonlinear oscillators with zero frequency 
($f=0$)
correspond to nonlinear trends.

\subsection{Model Equation}

We expand the given time series $S(t) \in \mathbb{R}$ with 
general complex functions $H_m(t) \in \mathbb{C}$ as 
\begin{equation}\label{eq-noe}
S(t) = \sum_{m=1}^M e^{H_m(t)},
\end{equation}
where $M$ is the number of the complex functions.

The complex functions are expressed as 
\begin{equation}\label{eq-h-fl}
H_m(t) = \ln c_m(t_0) + \int_{t_0}^t [ 2 \pi i f_m(\tau) + \lambda_m(\tau) ] {\rm d}\tau,
\end{equation}
where $f_m(t) \in \mathbb{R}$ represents the frequency modulation (FM) terms, 
which are known as ``instantaneous frequency'' by van der Pol~\cite{pol},
$\lambda_m(t) \in \mathbb{R}$ represents the amplitude modulation (AM) terms,
which correspond to our original \cite{cspa,isspit,ieice,thesis} nonlinear trends,
and $c_m(t_0) \in \mathbb{C}$ represents the amplitudes of the oscillators
at $t=t_0$.

This expansion corresponds to a mode decomposition with general complex functions,
noting that 
\begin{equation}
H_m^\prime(t) = 2 \pi i f_m(t) + \lambda_m(t).
\end{equation}

\subsection{Local Linearized Solution}

As it is known that our model equation Eq.~(\ref{eq-noe})
does not have a unique solution, 
due to the nonlinearity~\cite{daubeshies2011},
we need a special idea to make our model equation uniquely solvable.
For this purpose, we apply a local linearization technique 
\begin{equation}\label{eq-exp-h-local}
S(t)|_{t \sim t_k} \simeq \sum_{m=1}^M e^{H_m(t_k)+H_m^\prime(t_k) (t-t_k) + 
O\left((t-t_k \right)^2)}
\end{equation}
around $t \sim t_k=t_0+k \Delta T$,
consider a short enough time width, 
and ignore the higher order terms $O((t-t_k)^2)$ \cite{kubo}.

Then, 
the equation becomes a simple linear equation
\begin{equation}\label{eq-exp-h-locallinear}
S(t)|_{t \sim t_k} \simeq \sum_{m=1}^M e^{H_m(t_k)+H_m^\prime(t_k) (t-t_k) },
\end{equation}
and we can obtain unique $H_m^\prime(t_k)$ easily by using the 
linear predictive coding (LPC) method with $N$ samples,
noting that we must use a non-standard numerical method to solve LPC 
\cite{thesis,ccisp}.

Next, we calculate 
the complex amplitudes $c_m(t_k)$
of the oscillators
$c_m(t_k) e^{H_m^\prime(t_k) (t-t_k)}$ as
\begin{equation}\label{eq-calc-amps}
\argmin_{c_{m}(t_k)}
\sum_{n=0}^{N-1} 
\left(
S(t_k+n\Delta T)
-
\sum_{m=1}^M c_m(t_k) e^{n H_m^\prime(t_k) \Delta T}
\right)^2,
\end{equation}
and we obtain a local linearized solution
\begin{eqnarray}\label{eq-exp-h-locallinear-f-l}
S(t)|_{t \sim t_k} &\simeq& 
\sum_{m=1}^M c_m(t_k) 
e^{H_m^\prime(t_k)(t-t_k) }
\nonumber
\\
&=&
\sum_{m=1}^M c_m(t_k) 
e^{[2 \pi i f_m(t_k)+\lambda_m(t_k)] (t-t_k) }.
\end{eqnarray}

\subsection{Related Methods}

We list the major conventional methods in Table~\ref{tab-methods} \cite{thesis}.
In addition, we list the instantaneous frequency by van der Pol for reference.

FM terms $f_m(t)$ and AM terms $\lambda_m(t)$ for the conventional methods 
in the table 
are time-invariant,
meaning that they are linear methods.

\begin{table}[htbp]
\begin{center}
\caption{Comparison with major conventional methods.}
\label{tab-methods}
\centerline
{
\begin{tabular}{c|c|c|l}
Name &FM terms & AM terms & Comment
\\
\hline 
\hline
Our method & $f_m(t)$ & $\lambda_m(t)$ & 
calculates both functions
\\
\hline
\hline
van der Pol  & $f_\pm(t)$ & $0$ & 
$M=2$
\\
\hline
AR models
& $f_m$ & $\lambda_m < 0$ & time-invariant  T\oe plitz matrix 
\\
\hline
DFT &
$m/M \Delta T$ 
& 0 & prefixed time-invariant frequencies 
\\
\hline
STFT & 
$m/M \Delta T$ 
 & 0 & $c_m$ are stepwise time-variant
\\
\hline
\end{tabular}
}
\end{center}
\end{table}

The equation for 
the instantaneous frequency by van der Pol is
\begin{equation}
S(t)=\cos \int \omega(t) dt = 
\frac{1}{2}e^{i \int \omega(t) dt}+
\frac{1}{2}e^{-i \int \omega(t) dt},
\end{equation}
which corresponds to $M=2$.

Auto-Regressive (AR) models, such as LPC,
Maximum Entropy Method (MEM),
and MUltiple SIgnal Classification (MUSIC) \cite{music}
use a T\oe plitz matrix as an autocorrelation matrix, 
and time-invariant AM terms $\lambda_m(t)=\lambda_m$ 
are limited to negative values \cite{ccisp}.
In addition, they do not calculate complex amplitudes $c_m$.
Prony's method \cite{prony} is a kind of AR model,
which calculates $c_m$.
However, Prony's method is a historical one (published in 1795), 
and the method is considered only for reference.

Fourier transform methods such as the Discrete Fourier transform (DFT) 
and Short-Time Fourier Transform (STFT)
do not calculate FM terms $f_m(t)$, and they use prefixed 
time-invariant  frequencies $f_m(t)=m/M \Delta T$.
They calculate complex amplitudes 
$c_m$ of corresponding prefixed frequencies.
In addition, 
they ignore $\lambda_m(t)$, because they focus on periodic cases.


\section{Analysis}

\begin{figure}[hbt]
\begin{center}
\includegraphics[width=120mm]{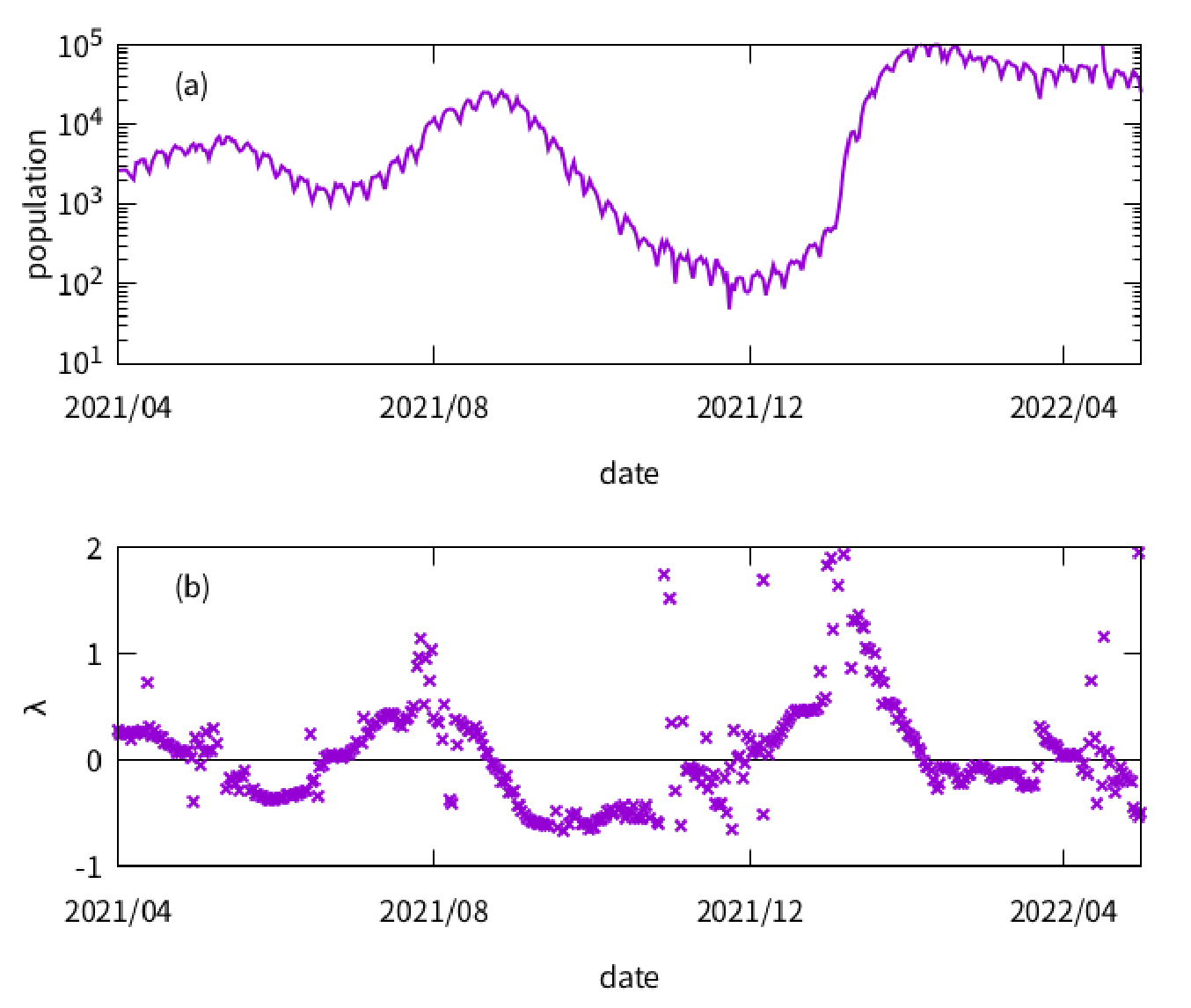}
\end{center}
\caption{
(a) Daily new cases in Japan,
and
(b) obtained nonlinear trends
in weekly basis.
}
\label{fig-jp}
\end{figure}

We used the WHO's COVID-19 global data \cite{who} for analysis,
and selected the time series $S(t)$ of daily new cases in Japan.
We show the time series in Fig.~\ref{fig-jp}(a).

Parameters for the analysis were set to $M=7,~N=14$.
That is, we used two weeks width data for a single analysis,
and we plotted the obtained nonlinear trends $\lambda$ 
in weekly basis $2^{\lambda t}$ as shown in Fig.~\ref{fig-jp}(b).
For example, $\lambda=1$ means that the number of infections doubles in a week.

Note that
major results are not affected by the parameters \cite{thesis,ccisp}.
Conventional methods are easily affected by the parameters,
because of the use of periodic boundary conditions.

In contrast to the 
case shown in Fig.~\ref{fig-cqg-b},
multiple nonlinear trends are not evident in Fig.~\ref{fig-jp}(b).
The system seems to have a single nonlinear trend, 
which holds for other countries also.
The reason for this will be explained in a further study.

\begin{figure}[hbt]
\begin{center}
\includegraphics[width=120mm]{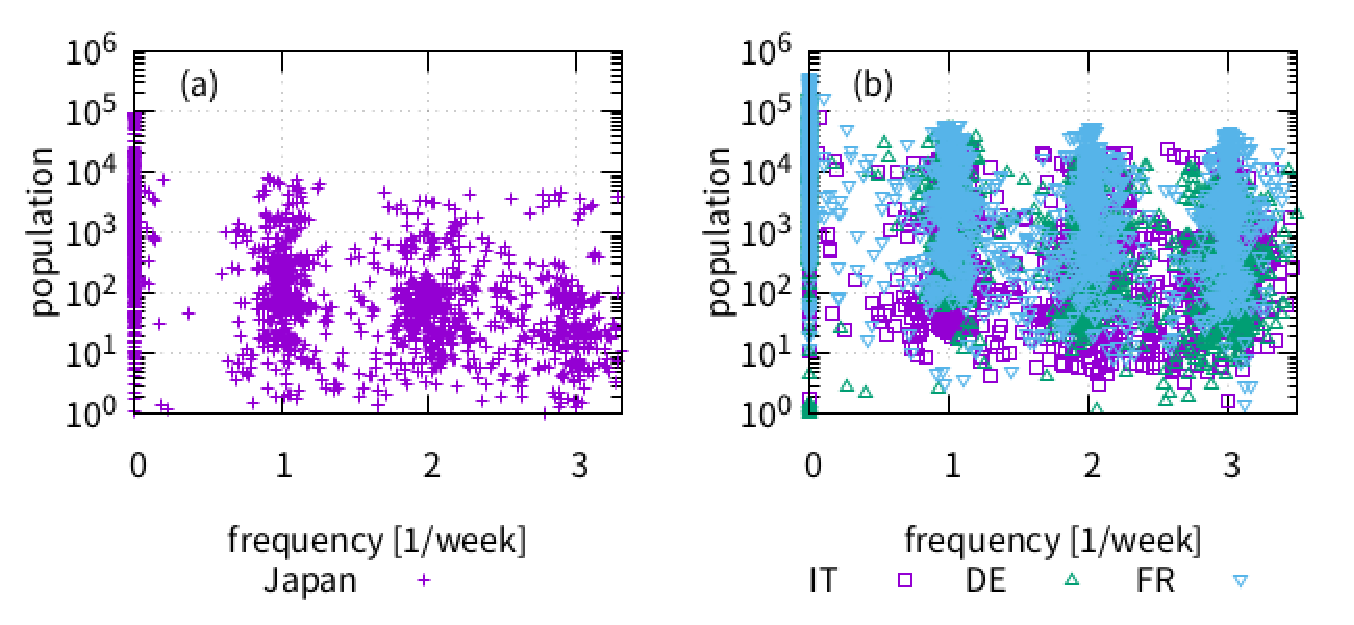}
\end{center}
\caption{
Spectra of (a) Japan, and (b) 
IT: Itary, DE: Germany, 
FR: France.
}
\label{fig-jp-p}
\end{figure}

We show
$|c_m(t_k)|$~vs~$f_m(t_k)$,
which corresponds to the spectrum, in Fig.~\ref{fig-jp-p}.
Spectra of several countries are shown.

The plotted time seres
in Fig.~\ref{fig-jp}(b) correspond to the plots
at 
the zero frequency region
$f=0$ 
in Fig.~\ref{fig-jp-p}(a).
Weekly oscillation $f=1$, 
and its second $f=2$ and third $f=3$ harmonics are also
plotted in Fig.~\ref{fig-jp-p}.
European countries in Fig.~\ref{fig-jp-p}(b) have much more clear harmonics.
As we focus on the nonlinear trends, these oscillation terms are 
not preferable.

As the system seems to have a single nonlinear trend, 
simplified analysis becomes possible.
That is, 
\begin{equation}\label{eq-ma}
\lambda(t)=\log_2 
\frac{\sum_{\tau=t-6}^{t} S(\tau)}{\sum_{\tau=t-13}^{t-7} S(\tau)},
\end{equation}
where weekly summation is used to suppress the weekly periodicity 
and its higher harmonics 
shown in Fig.~\ref{fig-jp-p}. 
This equation is known as the week-based infection growth rate,
and 
is justified only when there is a single nonlinear trend.

\begin{figure}[hbt]
\begin{center}
\includegraphics[width=120mm]{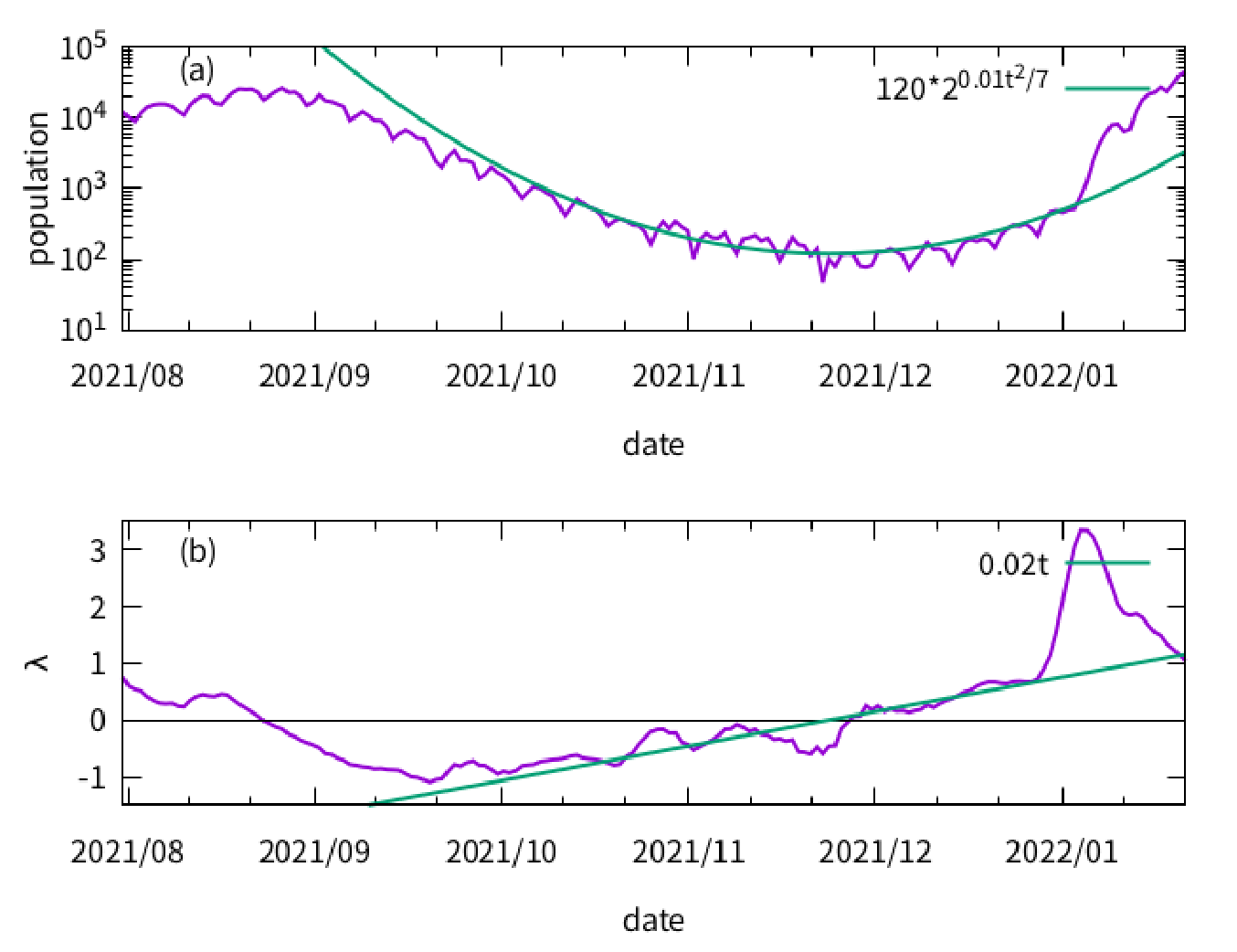}
\end{center}
\caption
{
(a) Daily new cases in Japan.
Line corresponds to Eq.~(\ref{eq-d-o-ts}) is also plotted.
(b) Obtained nonlinear trend
in weekly basis, using Eq.~(\ref{eq-ma}).
Line corresponds to Eq.~(\ref{eq-d-o-trend}) is also plotted.
}
\label{fig-d-o}
\end{figure}

We apply Eq.~(\ref{eq-ma})
to the Delta to Omicron variant transition in Japan,
as shown in Fig.~\ref{fig-d-o}.

The fitted line in Fig. \ref{fig-d-o}(b) is the analytical equation 
for the week-based nonlinear trend of the Delta variant
\begin{equation}\label{eq-d-o-trend}
\lambda(t)=0.02 (t-t_0),
\end{equation}
where $t_0$ is 24/Nov/2021,
and corresponding infection time series 
\begin{equation}\label{eq-d-o-ts}
S(t)=
120 \cdot 
2^{\int \lambda(t) dt/7}
=
120 \cdot 
2^{{0.01}(t-t_0)^2/7}
\end{equation}
is plotted in Fig.~\ref{fig-d-o}(a).
This fitted line holds from Sept/2021 to Jan/2022,
meaning that the infection status maintained the nonlinear trend
for over three months, 
and the transition to the Omicron variant happened in Jan/2022.
The Delta variant was replaced by the Omicron variant at that time,
and the extrapolated line in Fig.~\ref{fig-d-o}(b) after the replacement
will no longer hold.
Another fitting for the Omicron variant is required.

Obtained nonlinear time series Eq.~(\ref{eq-d-o-ts}) is explained as follows.
A time series $s(t)$ with 
a parameter 
$p$
and
with 
a lowest order perturbation term
$\epsilon  t$
becomes
\begin{equation}\label{eq-exp2}
s(t) \propto e^{\int_{0}^t p+\epsilon \tau d\tau}
=e^{pt+\frac{1}{2} \epsilon t^2}
= c e^{\frac{1}{2} \epsilon(t-t_0)^2},
\end{equation}
where $t_0$ is the time 
where $p+\epsilon t_0=0$,
and $c$ is the corresponding amplitude.
The linear term $p$ disappears, and the nonlinear term $\epsilon$ 
takes the place.


\section{Conclusion}

We demonstrated a nonlinear analysis
that provides an analytical perspective on a given numerical
time series.
We focused on the extraction of nonlinear trends
$\lambda_m(t) \propto t^{\alpha_m}$,
where $\alpha_m$ corresponds to the index of nonlinearity,
noting that
$\alpha_m=0$ is the linear case.

We applied our method 
to a time series of COVID-19 daily new cases in Japan,
and found that there is only a single nonlinear trend.
This result justifies the use of a week-based growth rate index.

The obtained nonlinear trend 
was
$\lambda(t) \propto t$,
and the fitting holds 
for
the Delta variant infection status
for a duration exceeding three months.
The fitting also visualized the transition to the Omicron variant.

The obtained nonlinear trend $\lambda(t)~\propto~t$ was also observed 
in another work \cite{lightning}, 
and this characteristics will be found widely,
because of the universality of Eq.~(\ref{eq-exp2}).

\appendix

\section{Expanding linear method into nonlinear method}

A linear system 
\begin{equation}
\frac{dx}{dt}- h x=0,
\end{equation}
where $h=2 \pi i f +\lambda$, 
has constant $h(t)=h$, and its corresponding time series 
becomes 
\begin{equation}
x=e^{\int h(t) dt}=e^{\int h dt}=e^{h t}.
\end{equation}
That is, non-constant  $h(t)$ means that the analyzed system is nonlinear.

A linear system with multiple $h_m$ is written as
\begin{equation}\label{eq-multi-lambda}
\left[
\prod_{m=1}^M \left( \frac{d}{dt}-h_m \right) 
\right]
 x=0,
\end{equation}
and its 
simple expansion 
into nonlinear form becomes
\begin{equation}\label{eq-multi-lambda-nonlin}
\left[
\prod_{m=1}^M \left( \frac{d}{dt}-h_m(t) \right) 
\right]
 x=0.
\end{equation}

However,  Eq.~(\ref{eq-multi-lambda-nonlin})
contains a problem.
That is, the equation is  affected by the order of the  multiplications
\begin{equation}\label{eq-incommutable}
\left[
\left( \frac{d}{dt}-h_1(t) \right) 
\left( \frac{d}{dt}-h_2(t) \right) 
\cdots
\left( \frac{d}{dt}-h_M(t) \right) 
\right]
 x=0,
\end{equation}
because of the existence of the term $d/dt$.

For example, the case $M=2$ becomes
\begin{eqnarray}
0&=&
\left[d/dt-h_1(t) \right] \left[d/dt-h_2(t) \right] x
\label{eq-app-m2}
\\
&=&
\left[d/dt-h_1(t) \right] \left[dx/dt-h_2(t) x \right] 
\nonumber
\\
&=&
d^2x/dt^2 - \underline{  d/dt \left[h_2(t) x \right] }
-h_1(t) \left[dx/dt-h_2(t) x \right] 
\nonumber
\\
&=&
d^2x/dt^2 - h_2(t) dx/dt - \underline{ h_2^\prime(t) x}
-h_1(t) dx/dt +h_1(t) h_2(t) x 
\nonumber
\\
&=&
d^2x/dt^2 -\left[ h_1(t)+h_2(t) \right]dx/dt 
+h_1(t) h_2(t) x - \underline{h_2^\prime(t) x}
\nonumber
\\
&=&
\left\{d^2/dt^2 -\left[ h_1(t)+h_2(t) \right]d/dt 
+h_1(t) h_2(t) \right\}x - \underline{h_2^\prime(t) x}.
\label{eq-nonlinear}
\end{eqnarray}

That is, an additional term $ - h_2^\prime(t) x $ appears, 
and this additional term has order dependency of the multiplications. 
For example, exchanging the  multiplication order
between $h_1(t)$ and $h_2(t)$
 in Eq.~(\ref{eq-app-m2}) gives 
an additional term $ - h_1^\prime(t) x $.

This order dependency is the very point of building nonlinear method of analysis.
A comuttable method is required to obtain a unique solution, 
and our method is the one.

We replace the functions $h_m(t)$ with corresponding constants $h_m(t_k)$
around
\begin{equation}
|t-t_k|<\epsilon,
\end{equation}
 and we obtain the linear equation
\begin{equation}\label{eq-multi-lambda-linearized}
\left[
\prod_{m=1}^M \left( \frac{d}{dt}-h_m(t_k) \right) 
\right]
 x_{t \sim t_k}=0
\end{equation}
to solve.

\section{Phenomenological boundaries of nonlinear trends}

\begin{figure}[hbt]
\begin{center}
\includegraphics[width=120mm]{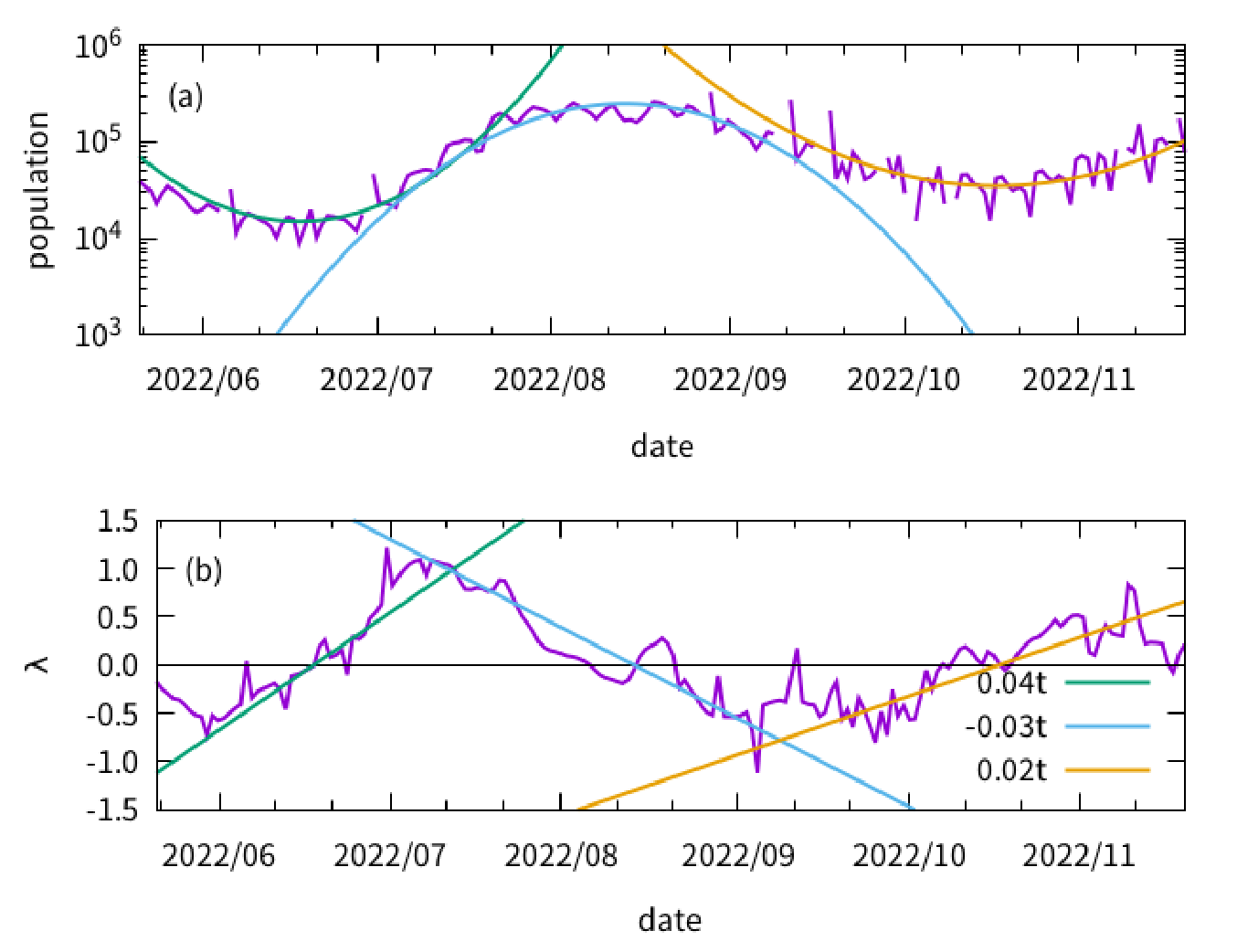}
\end{center}
\caption{
(a) Daily new cases in Japan.
Lines correspond to 
Eqs.~(\ref{eq-BS1}),(\ref{eq-BS2}),(\ref{eq-BS3}) are also plotted.
(b) Obtained nonlinear trend in weekly basis.
Lines correspond to 
Eqs.~(\ref{eq-BL1}),(\ref{eq-BL2}),(\ref{eq-BL3}) are also plotted.
}
\label{fig-appB}
\end{figure}

As the transition from the Delta to Omicron variant (Fig.~\ref{fig-d-o}) 
is an exceptional one,
we show some typical transitions in Fig.~\ref{fig-appB}.

We show the infection time series of 
Japan from 21/May/2022 to 20/Nov/2022
in Fig.~\ref{fig-appB}(a),
and its corresponding nonlinear trend in Fig.~\ref{fig-appB}(b).
Note that 
the reporting of all daily infection counts ended on 26/Sept/2022.
Since then, estimated number have been reported.

Fitted lines in Fig.~\ref{fig-appB}(b) are
\begin{eqnarray} 
\lambda_1(t) &=& ~~0.04 (t-t_1), \label{eq-BL1}\\
\lambda_2(t) &=& -0.03 (t-t_2),  \label{eq-BL2}\\
\lambda_3(t) &=& ~~0.02 (t-t_3), \label{eq-BL3}
\end{eqnarray}
where $t_1, t_2, t_3$ are  18/Jun, 14/Aug, 18/Oct, respectively.

And, their corresponding fitted time series in Fig.~\ref{fig-appB}(a) becomes
\begin{eqnarray}
S_1(t)&=& ~15000 \cdot 2^{{~~0.02~}(t-t_1)^2/7}, \label{eq-BS1}\\
S_2(t)&=&  250000 \cdot 2^{{-0.015}(t-t_2)^2/7}, \label{eq-BS2}\\
S_3(t)&=& ~35000 \cdot 2^{~~{0.01~}(t-t_3)^2/7}. \label{eq-BS3}
\end{eqnarray}

As our Eq.~(\ref{eq-exp2}) is based on the perturbation theory 
with the lowest order term,
which represents weak nonlinearity, 
it becomes less accurate as away from 
zero crossing points ($t_1$, $t_2$, $t_3$).
Therefore, three independent fittings were performed in Fig.~\ref{fig-appB}(b).

The crossing points of the fitted lines in Fig.~\ref{fig-appB}(b) 
are around $\lambda = \pm 1$, and they correspond to the phenomenological 
upper and lower boundaries, which structure is known as 
``the hammer and the dance'' \cite{pueyo}.

\begin{figure}[hbt]
\begin{center}
\includegraphics[width=120mm]{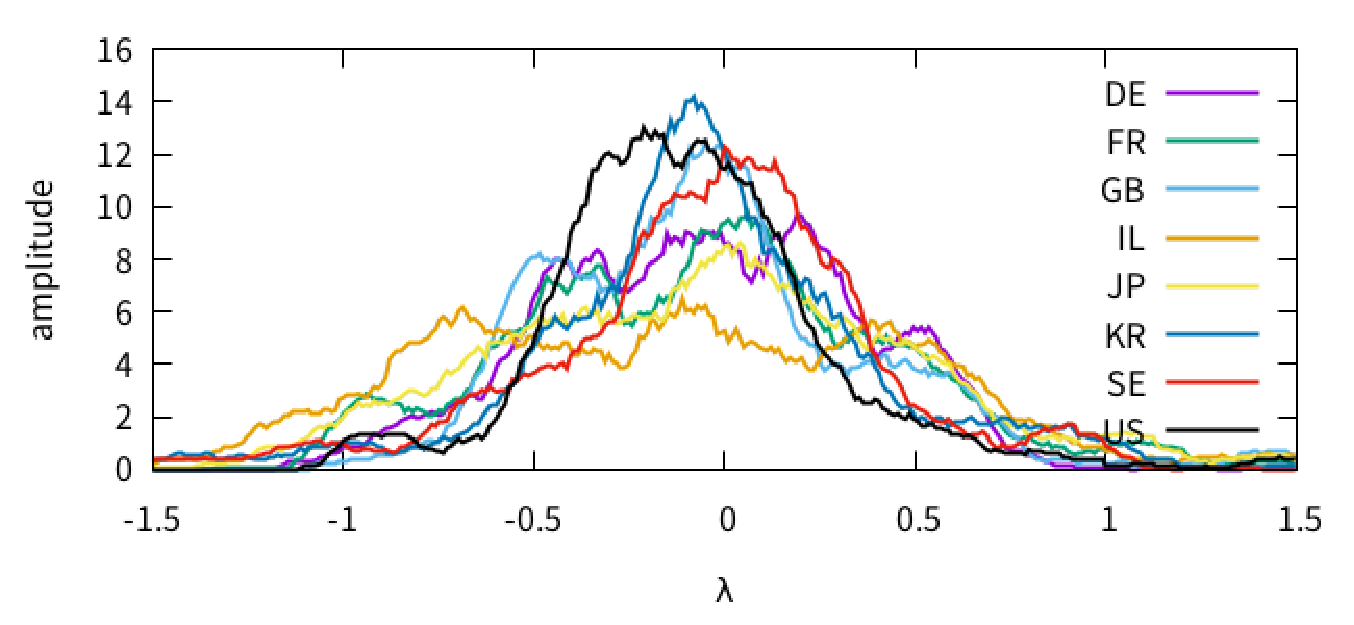}
\end{center}
\caption{
Distribution structures of nonlinear trend $\lambda$ in weekly basis.
DE:~Germany, FR:~France, GB:~England, IL:~Israel, 
JP:~Japan, KR:~Korea, SE:~Sweden, US:~United~States
}
\label{fig-appBhst}
\end{figure}

We plot the distribution structures of $\lambda$ in weekly basis 
for various countries in Fig.~\ref{fig-appBhst}.

We accumulated from 01/Mar/2020 to 28/Feb/2022,
and plotted the distributions in arbitrary units.

Among the countries,
Israel (IL, highly controlled with multiple vaccinations) and 
Sweden (SE, without control) are the typical countries.

The phenomenological boundaries $\lambda = \pm 1$ seem to 
hold for the countries shown in the figure,
even though most of the countries tried hard lockdown.
Some peak structures should appear in $\lambda<-1$,
but no peak is observed in the figure.
Instead, several countries (DE, FR, GB) show peak structure around $\lambda=-0.5$,
and the structures are a candidate for lockdown.



\begin{thebibliography}{11}

\bibitem{ieee-access}
F. Rustam et al., 
``COVID-19 future forecasting using supervised machine learning models,''
{\it
IEEE Access,
}
vol. 8, pp. 101489-101499, 2020.
\\
DOI: 10.1109/ACCESS.2020.2997311

\bibitem{ieee-s}
S. Vanahalli and P. N, 
``An intelligent system to forecast COVID-19 pandemic 
using hybrid neural network,'' 
{\it
Proc.
10th IEEE Int. Conf. Commun. Syst. Netw. Tech. (CSNT),
}
pp. 543-548, 2021.
\\
DOI: 10.1109/CSNT51715.2021.9509622

\bibitem{ieee-y}
Y. Alali, F. Harrou and Y. Sun, 
``Optimized Gaussian process regression by Bayesian optimization to 
forecast COVID-19 spread in India and Brazil: a comparative study,'' 
{\it
Proc.
Int. Conf. ICT Smart Soc. (ICISS), 
}
pp. 1-6, 2021.
\\
DOI: 10.1109/ICISS53185.2021.9532501


\bibitem{cspa}
F. Ishiyama, Y. Okugawa and K. Takaya,
``Linear predictive coding without Yule-Walker approximation 
for transient signal analysis - application to switching noise,''
{\it 
Proc. 13th IEEE Colloq. Signal Process. Appl., 
}
pp. 46-50, 2017.
\\ 
DOI: 10.1109/CSPA.2017.8064922

\bibitem{isspit}
F. Ishiyama,
``Local linear predictive coding for high resolution time-frequency analysis,''
{\it
Proc. 17th IEEE Int. Symp. Signal Process. Info. Tech. (ISSPIT), 
}
pp.~1-6, 2017.
\\ 
DOI: 10.1109/ISSPIT.2017.8388309

\bibitem{ieice}
F. Ishiyama,
``Piecewise linear predictive coding for nonlinear signal analysis 
and automatic trend extraction,''
{\it IEICE Trans. Fundamentals (Japanese Edition)}, 
vol. J101-A, pp. 36-45, 2018.

\bibitem{thesis}
F. Ishiyama,
``Constructing nonlinear signal processing theory for 
characterization and identification of electromagnetic noise sources,
and its feasibility study,''
Ph.~D. Thesis,
\emph
{Univ. of Tsukuba}, 2021.
\\
DOI: 10.15068/0002000757

\bibitem{git}
Sample codes for our method:
\\

\bibitem{cqg}
F. Ishiyama, and R. Takahashi,
``The bounce hardness index of gravitational waves,'' 
{\it
Class. Quant. Grav.,
}
vol. 27, 245021 (11pp), 
2010.
DOI: 10.1088/0264-9381/27/24/245021

\bibitem{dimm}
H. Dimmelmeier, J. A. Font and E. M\"{u}ller,
``Relativistic simulations of rotational core collapse II. Collapse dynamics and gravitational radiation,'' 
{\it
Astron. Astrophys.,
}
vol. 393, pp. 523-542, 2002.
\\
DOI: 10.1051/0004-6361:20021053

\bibitem{pol}
B. van der Pol,
``The fundamental principles of frequency modulation,''
{\it
J. Inst. Elect. Eng. III, 
}
vol. 93, pp. 153-158, 1946.

\bibitem{daubeshies2011}
I. Daubechies, J. Lu and H. T. Wu,
``Synchrosqueezed wavelet transforms: An empirical mode decomposition-like tool,''
{\it Appl. Comput. Harmon. Anal.}, 
vol. 30, pp. 243-261, 2011.

\bibitem{kubo}
R. Kubo,
``Statistical-mechanical theory of irreversible process. I.
 General theory and simple applications to magnetic and conduction problems,''
{\it 
J. Phys. Soc. Jpn., 
}
vol. 12, pp. 570-586, 1957.
\\
DOI: 10.1143/JPSJ.12.570

\bibitem{ccisp}
F. Ishiyama, 
``Maximum entropy method without false peaks with exact numerical equation,'' 
{\it 
J. Phys.: Conf. Ser., 
}
vol. 1438, 012031 (6pp), 2020.
\\ 
DOI: 10.1088/1742-6596/1438/1/012031

\bibitem{music}
R. O. Schmidt,
``Multiple Emitter Location and Signal Parameter Estimation,''
{\it IEEE Trans. Antennas Propag}., vol. AP-34, pp. 276-280, 1986. 

\bibitem{prony}
R. Prony,
``Essai exp\'erimental et analytique - 
Sur les lois de la Dilatabilit\'e des fluides \'elastiques 
et sur celles de la Force expansive de la vapeur de l'eau et de la vapeur de l'alkool,
\`a diff\'erentes temp\'eratures,'' 
{\it J. l'\'Ecole Polytechnique},
vol. 1,  Flor\'eal et Plairial III, 
pp. 24-76, 1795.

\bibitem{who}

\bibitem{lightning}
F. Ishiyama and M. Maruyama, 
``Nonlinear time-frequency analysis of lightning strike surge current waveforms
recorded at Gasing hill, Kuala Lumpur,'' 
{\it
Proc. 18th IEEE Int. Colloq. Signal Process. Appl. (CSPA),
}
pp. 20-23, 2022.
\\ 
DOI: 10.1109/CSPA55076.2022.9782060

\bibitem{pueyo}

\end{thebibliography}
\end{document}